\documentclass[11pt,reqno]{amsart}
\def\@rmrk#1#2{\refstepcounter
    {#1}\@ifnextchar[{\@yrmrk{#1}{#2}}{\@xrmrk{#1}{#2}}}

\newfont{\bfit}{cmbxti10 scaled 2000}
 \newfont{\biggi}{cmr12 scaled 2000}

 \newcommand{\eps}{\varepsilon}

 \newcommand{\R}{\mathbb{R}}
 
 \newcommand{\N}{\mathbb{N}}

 \newcommand{\me}{\mathbb{E}}

 \newcommand{\one}{\1}

 \newcommand{\cl}{{\rm cl}}
 \newcommand{\interior}{{\rm int}}

 \newcommand{\tome}{\tilde{\omega}}

 \newcommand{\skric}{{\mathcal C}}

 \newcommand{\skrie}{{\mathcal E}}

 \newcommand{\skril}{{\mathcal L}}
 \newcommand{\skrim}{{\mathcal M}}

 \newcommand{\skriv}{{\mathcal V}}
 \newcommand{\skrix}{{\mathcal X}}
 \newcommand{\skriy}{{\mathcal Y}}
 
 \newcommand{\heap}[2]{\genfrac{}{}{0pt}{}{#1}{#2}}
 \newcommand{\sfrac}[2]{\mbox{$\frac{#1}{#2}$}}

\def\1{{\mathchoice {1\mskip-4mu\mathrm l}      
{1\mskip-4mu\mathrm l}
{1\mskip-4.5mu\mathrm l} {1\mskip-5mu\mathrm l}}}

\newcommand{\eq}{\begin{equation}}
\newcommand{\en}{\end{equation}}

\newenvironment{Proof}
{\vskip0.1cm\noindent{\bf Proof. }{\hspace*{0.3cm}}}%
{\nopagebreak {\hspace*{\fill}\rule{2mm}{2mm}}\\ }
{\nopagebreak {\hspace*{\fill}\rule{2mm}{2mm}}\\ }

\renewcommand{\subsection}{\secdef \subsct\sbsect}
\newcommand{\subsct}[2][default]{\refstepcounter{subsection}
\vspace{0.15cm}
{\flushleft\bf \arabic{section}.\arabic{subsection}~\bf #1  }
\nopagebreak\nopagebreak}
\newcommand{\sbsect}[1]{\vspace{0.1cm}\noindent
{\bf #1}\vspace{0.1cm}}

\newtheorem{theorem}{Theorem}[section]
\newtheorem{lemma}[theorem]{Lemma}
\newtheorem{cor}[theorem]{Corollary}

\newtheoremstyle{thm}{1.5ex}{1.5ex}{\itshape\rmfamily}{}
{\bfseries\rmfamily}{}{2ex}{}

\newtheoremstyle{rem}{1.3ex}{1.3ex}{\rmfamily}{}
{\itshape\rmfamily}{}{1.5ex}{}
\theoremstyle{rem}

\refstepcounter{subsection}

\def\thebibliography#1{\section*{References}
  \list%
  {\arabic{enumi}.}
    {\settowidth\labelwidth{[#1]}\leftmargin\labelwidth
    \advance\leftmargin\labelsep
    \parsep0pt\itemsep0pt
    \usecounter{enumi}}
    \def\newblock{\hskip .11em plus .33em minus .07em}
    \sloppy                   
    \sfcode`\.=1000\relax}

\begin{document}
\title[Random partition function  for Ising  model on  spinned random  graphs] {Asymptotics  of  the partition function  of  Ising  model on  Inhomogeneous random  graphs}
\author[K.D-A]{}

\maketitle
\thispagestyle{empty}
\vspace{-0.5cm}

\centerline{\sc{Kwabena Doku-Amponsah}}

\vspace{0.5cm}

 \centerline{\bf \small Abstract}
\begin{quote}{\small For a finite  random graph,  we  defined a  simple model  of  statistical  mechanics. We  obtain an annealed  asymptotic result for  the  random partition function for this model on finite random  graphs  as $n,$ the  size  of  the graph  is  very large. To  obtain this  result, we define  the \emph{ empirical bond distribution}, which  enumerates the  number  of  bonds between a  given couple of spins,  and \emph{ empirical spin distribution},   which  enumerates the  number of sites having a given spin  on the  spinned  random  graphs.  For  these  empirical distributions  we  extend  the  large  deviation  principle(LDP)   to cover random graphs with continuous colour laws. Applying  Varandhan Lemma  and this  LDP   to  the  Hamiltonian of the Ising model defined on  Erdos-Renyi graphs, expressed as  a  function  of  the empirical distributions,  we  obtain our annealed asymptotic result.}
\end{quote} \vspace{0.5cm}


{\textit{Mathematics Subject Classification :} 46N55,  58Z05,  60F10,05C80.\\

\section{Introduction}
In  this  paper  we  study  the  partition  function  of  the  Ising  model  on  an  inhomogeneous  random  graph  model  defined as  follows: we  first  divide vertices  into two  types,  where  the  spin  of  every  node  is chosen independently  and  identically  according  to some  spin  law,  then each  bond is  drawn  independently  with  its  probability  given  depending on  the  spins  of  the  endpoints.This  graph model is  similar  to  the  planted partition  or  stochastic  block model, see \cite{1},  with  the slight  difference  that  in  this  model the  edge  probability  within a  block(type)  is  allowed  to be  different  between  the  two  blocks. On  this  random graph,  we   consider  the  Ising  model  and  establish  the  asymptotic  limit  of  the  normalized  log  of  the  expected  partition  function,  where  the  expectation  is  over  the  random  graph.

The  random  partition  function  of  random  graphs  has  generated  considerable  interest  in  the  mathematical physics  community, see examples \cite{2}, \cite{3},\cite{4},\cite{5},\cite{6},\cite{7}, \cite{8},\cite{9},\cite{10},\cite{11}, \cite{12}, \cite{13}, \cite{14}, for  the understanding   of  the  log  partition  function of  constraint satisfaction  problems(such as  the  Ising/Potts  models,  legal  colourings,  hardcore  model  etc.)  on  a  random  graph. 
See, \cite{15}, \cite{16}, \cite{17}, \cite{18}  for  some recent  applications  of  the Ising  model  on  random  graphs.

 The  most  sought after  quantity, which may  be  quite challenging to  analyse  in  various  models,  is  the  quenched  setting. The  log-partition  function  is typically not concentrated  and  the  analyses of  the quenched log-partition is  only possible  for locally tree-like  random graphs. \cite{2},\cite{8},\cite{10},\cite{11}, \cite{12}, \cite{13}  provide examples  of   the quenched asymptotic results for the  log-partition  function  for  the  Ising model  on  general  local tree-like  shape  random graphs which inhomogeneous  Curie-Weiss  and  annealed random  graphs. In  \cite{5}, an  annealed asymptotic result for the  random partition  function was  first proved   and the analogue  of  the  quenched  results  were  proved  for  the  Potts  model  on  a  regular  graph.

The  proofs  in  all these  papers  rely  on  the general  locally homogeneous tree-like  shape property of  the random  graphs; the configuration model, planted  partition  random  graph model, stochastic block model. Some researchers, see e.g \cite{14},  have use the Monte Carlo simulations to  investigate  the asymptotic properties of  the Ising model defined on Erdos-Renyi graphs.\\

 In this  paper we  study statistical mechanic  on random graphs  which  are  \emph{locally non-homogeneous tree-like}. Thus  we  study random  graphs  with  a local structure of multitype  Galton-Watson branching trees.  To  be  specific, we use the large  deviation  principle (LDP) techniques developed in \cite{19}  and furthered in \cite{20} to prove  annealed  asymptotic  result for  the log-partition functions  of  Ising  model on  inhomogeneous random graphs. Our  annealed asymptotic result may  serve as  the  basis for understanding  the  thermodynamic  limiting  behaviour  of  the  free-energy  function  of  the  Ising  model  on  inhomogeneous  graphs.\\

The  remaining  part  of  the  article is  organized as follows: We present in Section~\ref{section2} the  Ising  model  on  spinned  graphs. The main results of  this  article, Theorem~\ref{mainpart}   and  a  Corollary~\ref{mainpart1}  are stated in  Section~\ref{main}. Section~\ref{LDP} contains the  LDP  result   and  proof. Finally,  we use  our  LDP  results  to  prove  our  main  Theorem  and  derive  Corollary  in Section~\ref{mainproof}.

\section{The Model}\label{section2}
By $\skriv=\{1,2,3,...,n\}$  we denote by fixed set of $n$  sites and
$E\subset\skrie:=\big\{(v,u)\in \skriv\times \skriv \, : \, v<u\big\},$ where
the \emph{formal} ordering of bonds  is just  to  help us to simply describe \emph{unordered} links.
Let $\skrix=\{-1,+1\}$ be the spin set and  define  a  function $\rho$  by  $\rho:\skriv\to\skrix.$

For a symmetric function
$q_n\colon\R\times\R\rightarrow [0,1],$ a continuous function
$\sigma:\skrix\rightarrow\R$  and a probability distribution $\ell$ on
$\sigma(\skrix)=\big\{\sigma(-1),\,\sigma(+1)\big\}$ the {\em Inhomogeneous spinned random
graph} or\emph{ spinned  graph}~$G$ with $n$ sites may  be  obtained
as follows:

Each site $u\in \skriv$  with $\rho(u)\in\skrix$  is  assigned \emph{magnetic ising spin }$\sigma(\rho(u))$
independently according to the {\em spin law} $\ell.$ Given the
spins, we link any two sites $v,u\in \skriv$, independently according to a {\em link probability}
$q_n[\sigma(\rho(v)),\sigma(\rho(u))]$ otherwise they remain disconnected.

We always observe  $G=\{\big(\sigma(\rho(v))\,:\,v\in \skriv\big),E\}$ under
the joint probability measure of graph and spin. We shall call $G$  spinned
graph and observe $\sigma(\rho(v))$ as the spin of the site $v$.\\

On a spinned graph $X,$  we define  the ferromagnetic  Ising Model
by the following \emph{Boltzmann  distributions}  over $\skrix^\skriv$ ,
$$\mu_G(\rho)=\sfrac{1}{Z_G(\beta,B)}\exp\Big\{\sum_{(v,u)\in E}\sfrac{\sigma(\rho(u))}{B(\rho(u))}\sfrac{\sigma(\rho(v))}{B(\rho(v))}
+\sum_{u\in \skriv}\sfrac{\sigma(\rho(u))}{\sqrt{\beta}}\Big\}$$ where
$\sigma(\rho(u))=\sqrt{\beta}B(\rho(u))\rho(u),$  $\beta\ge 0$  is  the
inverse temperature, $B=\{b(x): x\in \{-1,+1\}^n\}$ is the vector
external of magnetic fields, $\rho(u)\in\{-1,+1\},$ and $Z_G $ the
random partition function  is  given by
$$Z_G(\beta,B):=\sum_{\rho\in\{-1,+1\}^\skriv}\exp\Big\{\sum_{(v,u)\in
E}\sfrac{\sigma(\rho(u))}{B(\rho(u))}\sfrac{\sigma(\rho(v))}{B(\rho(v))}+\sum_{u\in
V}\sfrac{\sigma(\rho(u))}{\sqrt{\beta}}\Big\}.$$
 Note $\sfrac{\sigma(\rho(v))}{B(\rho(v))}=0$  if  $B(\rho(v))=0$  and  also  $\sfrac{\sigma(\rho(v))}{\sqrt{\beta}}=0$  if $\beta=0.$

Our  main concern  in this paper is  the  study  of  the  annealed  asymptotic of the partition function  of the  Ising  model  of  spinned  random graphs.  We assume that $X$  is  near-critical .i.e. the bond probabilities satisfy:
$$\lim_{n\to \infty}n q_n[\sigma(+1),\sigma(+1)] = C[\sigma(+1),\sigma(+1)],\lim_{n\to \infty}n q_n[\sigma(-1),\sigma(+1)] =
C[\sigma(-1),\sigma(+1)]$$   $$\lim_{n\to \infty}n q_n[\sigma(+1),\sigma(-1)] = C[\sigma(+1),\sigma(-1)],
\lim_{n\to \infty}nq_n[\sigma(-1),\sigma(-1)] = C[\sigma(-1),\sigma(-1)],$$
where $C\colon\R\times\R\rightarrow [0,\infty)$ is a symmetric
function and is not equal to zero identically.  Throughout  the  rest part  of  the  article  we  will  write $$C_{i,j}=C[(\sigma(i),\sigma(j)].$$

\section{Main Result}\label{main}
We  define  $\lambda:[0,\,1]\to  \R$  a  function  by

$$\lambda(y)=a_1 y^2+ a_2(1-y)^2+a_3y(1-y)+B(1)y-B(-1)(1-y),$$
where
$$ a_1(\beta,\, B)=\sfrac12\,C_{+1,+1}(e^{\beta}-1),\, a_2(\beta,\, B)=\sfrac12\,C_{-1,-1}(e^{\beta}-1) \, \mbox{and}\,\, a_3(\beta,\, B)=\,C_{-1,+1}(e^{-\beta}-1).
$$
 We recall that  $B$ is  a  vector  of  magnetic  field of  the  system  and state the  main  result of  this  article,   Theorem~\ref{mainpart}.

\begin{theorem}\label{Ising.1}\label{main1}\label{mainpart}
Suppose that $X$ is a spinned random graph with bond probabilities $
q_n\colon\skrix\times\skrix\rightarrow[0,1]$ satisfying $nq_n[\sigma(e),\sigma(t)] \to C[\sigma(e),\sigma(t)], $  for all  $t,e\in \{-1,+1\}$ and
some continuous symmetric function $C\colon\R\times\R\rightarrow
[0,\infty)$ with
$\sup_{y,x\in\R}C[y,\,x]<\infty.$  Let $Z_G(B,\beta)$ be  random  partition  function  of   $X.$
  Then,  $Z_G(B,\beta)$   satisfies
$$\lim_{n\to\infty}\frac{1}{n}\log\me\Big [ Z_G(B,\beta)\Big]=\sup_{y\in[0,\,1]}\Big\{-y\log y-(1-y)\log(1-y)+\lambda(y)\Big\}.$$

\end{theorem}

From  Theorem~\ref{mainpart}  we  derive a  corollary for the  log- partition function of   the  Erdos-Renyi  Graphs (where  $np_n\to c $) as  a  special  case  of  random  graphs  with locally non-homogeneous tree-like  property.

\begin{cor}\label{Ising.2}\label{main2}\label{mainpart1}
Suppose that $X$ is an  Erdos-Renyi random graph with bond probability such  that  $nq_n \to c.$  Let $Z_G(B,\beta)$ be  random  partition  function  of   $X.$  Then, $Z_G(B,\beta)$  satisfies
\begin{equation}\begin{aligned}\label{mainErdos-Renyi}
\lim_{n\to\infty}\frac{1}{n}\log\me\Big [ Z_G(B,\beta)\Big]=\sup_{y\in[0,\,1]}\Big\{-y\log y-(1-y)\log(1&-y)+c(e^{\beta}-1)(y^2-y+\sfrac12)\\
&+c(e^{-\beta}-1)(y-y^2)+2by-b\Big\},
\end{aligned}
\end{equation}
where $B(+1)=B(-1)=b$  is  constant external  magnetic field.

\end{cor}

\section{Large deviation principles   for spinned random  graphs}\label{LDP}

In this section, we  review some large deviation  results of
 \cite{19},  and  extend  the  joint  LDP  for
empirical  distributions of  coloured random graph  to  spinned random
graphs.i.e. we  assume a  more general  spin law $\ell:\R\to [0,1]$
with  all its  exponential moments finite  and prove an  LDP  for
this  model in  a  topology generated by the total  variation norm.

To  begin,  we recall some useful  definitions and notations from
\cite{19}.  A \emph{rate function} is a
non-constant, lower semi-continuous function $J$ from a polish space
$\skril$ into $[0,\infty]$, we called  it \emph{good} if the level
sets $\{ J(m)\le \alpha\}$ are compact for every $\alpha\in [0,\infty)$. A
functional $L$ from the set of finite spinned  random graph model to $\skril$ is
said to satisfy a \emph{large deviation principle} with rate
function $J$ if, for all Borel sets $\Gamma\subset \skril$,
\begin{equation}\nonumber
\begin{aligned}
-\inf_{p\in \interior\,\Omega} J(p)  \le \liminf_{n \to \infty} \sfrac 1n \log P\big\{ L(X) \in \Omega \big\}
 \le \limsup_{n \to \infty} \sfrac 1n \log P\big\{ L(X) \in \Omega
\big\} \le -\inf_{p\in \cl\,\Omega} J(p)\, ,
\end{aligned}
\end{equation}

where $X$ under $P$ is a spinned random graph with $n$
sites and int $\Omega$ and cl $\Omega$ refer to the interior, resp.
closure, of the set $\Omega$.

And, for any finite or countable set $\skriy$ we denote by
$\skril(\skriy)$ the space of probability distributions, and by
$\tilde\skril(\skriy)$ the space of finite distributions on $\skriy$,
both endowed with the topology   generated   by  the total variation
norm. For $\omega\in\tilde\skril(\skriy)$ we denote by $\|\omega\|$
its total mass. Further, if $\ell\in\tilde\skril(\skriy)$ and
$\nu\ll\ell$ we denote by
$$H(\eta\,\|\,\ell)=\int_{\R}\eta[dy]
\log\big(\sfrac{\eta[dy]}{\ell[dy]}\big)$$ the \emph{relative
entropy} of $\eta$ with respect to $\ell$. We set
$H(\eta\,\|\,\ell)=\infty$ if
$\eta\not\ll\ell$. 
Finally, we denote by $\tilde\skril_*(\skriy \times \skriy)$ the
subspace of symmetric distributions in $\tilde\skril(\skriy \times
\skriy)$.\\

On each spinned graph $G=((\sigma(\rho(v))\,:\,u\in \skriv),E)$ with $n$
vertices, we define  a probability distribution, the \emph{empirical spin
distribution}~$\skrim_G^{1}\in\skril(\skrix)$,~by
$$\skrim_G^{1}[y]:=\frac{1}{n}\sum_{u\in \skriv}\delta_{\sigma(\rho(u))}(y),\quad\mbox{ for $y\in\R$, }$$
and a symmetric finite distribution, the \emph{empirical bond
distribution} $\skrim_G^{2}\in\tilde\skril_*(\R\times\R),$ by
$$\skrim_G^{2}[y,x]:=\frac{1}{n}\sum_{(v,u)\in E}[\delta_{(\sigma(\rho(v),\,\sigma(\rho(u))}+
\delta_{(\sigma(\rho(u)),\,\sigma(\rho(v))}](y,x),\quad\mbox{ for
$x,y\in\R$. }$$ The total mass $\|\skrim_G^{2}\|=2|E|/n$.

\begin{theorem}[Extension]\label{Ising.4}\label{main4}
Suppose that $G$ is a spinned random graph with spin law $\ell$ such
 that  $n^{-1} \log \ell(n) \to -\infty$   and bond probabilities  $ q_n\colon\R\times\R\rightarrow[0,1]$
satisfying $n q_n[dy,dx] \to C[dy,dx], $  for all  $x,y\in \R$ and
some bounded symmetric distribution
$C\colon\R\times\R\rightarrow [0,\infty).$  
 Then, as  $n\rightarrow\infty,$ the pair $(\skrim_G^1,\skrim_G^2)$ satisfies  a
large deviation principle in
$\skril(\R)\times\tilde{\skril}_*(\R\times\R)$ with good rate
function
\begin{equation}\label{randomg.rateL2L1}
J(\eta,\varpi)=H(\eta\,\|\,\ell)+\frac{1}{2}\Big[H\big(\omega\,\|\,C\eta\otimes\eta\big)+\|
C\eta\otimes\eta \| -\|\omega\|\Big]\,.
\end{equation}

     where the distribution
$C\eta\otimes\eta\in\tilde\skril(\R\times\R)$ is defined by
$C\eta\otimes\eta[dy,dx]=C(x,y)\eta[dx]\eta[dy]$.
\end{theorem}

\section{Proof of  Large Deviation Theorem  Extension}\label{extension}
\subsection{Exponential Change-of-Measure}
 Given a bounded function $\tilde{f}\colon\skrix\rightarrow\R$ and
a symmetric bounded function $\tilde{g}\colon
\skrix\times\skrix\rightarrow\R$,
we define the constant $U_{\tilde{f}}$ by
$$U_{\tilde{f}}=\log\int_{\R}e^{\tilde{f}[x]}\ell[dx],$$
and the function $\tilde{h}_n\colon\R\times\R\rightarrow\R$ by
\begin{equation}\label{hdef}
\tilde{h}_n[x,y]=\log\Big[\big(1-q_n[dy,dx]+ q_n[dy,dx]
e^{\tilde{g}[x,y]}\big)^{-n}\Big],
\end{equation}
for $a,b\in\skrix.$ We use $\tilde{f}$ and $\tilde{g}$ to define
(for very large $n$) a new spinned random graph  as
follows:
\begin{itemize}
\item To the $n$ labelled sites in $\skriv$ we assign spins from
$s(\skrix)$ iid according to the spin
law $\ell$ defined by
$$ \tilde{\ell}[dx]=e^{\tilde{f}[x]-U_{\tilde{f}}}\ell[dx].$$
\item Given  any two sites $v,u\in \skriv,$ with $u$ carrying spin $x$ and
$v$ carrying spin $y$, we link site  $u$ to site $v$ with
probability
$$\tilde{q}_{n}[y,x]=\frac{q_n[dy,dx]e^{\tilde{g}[y,x]}}{(1-q_{n}[y,x]+q_n[dy,dx])e^{\tilde{g}[y,x]}}
.$$
\end{itemize}
We denote the transformed law by $\tilde{P}.$  We observe that
$\tilde{\ell}$ is a probability distribution and that  $\tilde{P}$ is
absolutely continuous with respect to $P$ as, for any coloured
graph $G=((\sigma(\rho(u))\colon u\in \skriv),E)$,
\begin{align}
\frac{d\tilde P}{d P}(G) & = \prod_{u\in
V}\sfrac{\tilde{\ell}[\sigma(\rho(u))]}{\ell[\sigma(\rho(u))]}\prod_{(v,u)\in
E}\sfrac{\tilde{q}_{n}[\sigma(\rho(v)),\sigma(\rho(u))]}
{q_{n}[\sigma(\rho(v)),\sigma(\rho(u))]}\prod_{(v,u)\not\in E}\sfrac{1-\tilde{q}_{n}[\sigma(\rho(v)),\sigma(\rho(u))]}{1-q_{n}[\sigma(\rho(v)),\sigma(\rho(u))]}\nonumber\\
& = \prod_{u\in
V}\sfrac{\tilde{\ell}[\sigma(\rho(u))]}{\ell[\sigma(\rho(u))]}\prod_{(v,u)\in
E}\sfrac{\tilde{q}_{n}[\sigma(\rho(v)),\sigma(\rho(u))]}{q_{n}[\sigma(\rho(v)),\sigma(\rho(u))]}
\times\sfrac{n-nq_{n}[\sigma(\rho(v)),\sigma(\rho(u))]}{n-n\tilde{q}_{n}[\sigma(\rho(v)),\sigma(\rho(u))]}\prod_{(v,u)\in\skrie}\sfrac{n-n\tilde{q}_{n}[\sigma(\rho(v)),\sigma(\rho(u))]}
{n-nq_{n}[\sigma(\rho(v)),\sigma(\rho(u))]}\nonumber\\
& = \prod_{u\in
V}e^{\tilde{f}[\sigma(\rho(u))]-U_{\tilde{f}}}\prod_{(v,u)\in
E}e^{\tilde{g}[\sigma(\rho(v)),\sigma(\rho(u))]}\prod_{(v,u)\in\skrie}{e^{\frac 1n\, \tilde{h}_n[\sigma(\rho(v)),\sigma(\rho(u))]}}\nonumber\\
& = \exp\big( n\langle \skrim_G^1, \tilde{f}-U_{\tilde{f}}\rangle+n\langle
\sfrac{1}{2}\skrim_G^2, \tilde{g}\rangle+n\langle\sfrac{1}{2}\skrim_G^1\otimes
\skrim_G^1, \tilde{h}_n\rangle-\langle \sfrac{1}{2}\skrim_{\Delta}^{1},
\tilde{h}_n\rangle \big) , \label{Ising.Itransform}
\end{align}
where  $$\skrim_{\Delta}^{1}=\sfrac{1}{n}\sum_{v\in
V}\delta_{(\sigma(\rho(v)),\sigma(\rho(v)))}.$$

We write  $\langle f,\eta\rangle:=\int_{\R} f[x]\eta[dx]$ and
$$\langle g,\omega\rangle:=\int_{\R\times\R} g[y,x]\omega[dy,dx].$$

\begin{lemma}[Euler's lemma]\label{Ising.sequence1}
If $nq_n[dy,dx]\to C[dy,dx]$ for every $y,x\in\R$, then
\begin{equation}\label{Ising.sequence}
\lim_{n\rightarrow\infty}\Big(1+\alpha q_{n}[y,x]\Big)^{n}=e^{r
C[dy,dx]},\mbox {for all $y,x\in\R$  and for  $r\in\R$.}
\end{equation}
\end{lemma}

\begin{Proof}
Notice that, for any $\eps>0$ and for very large $n$  we have
$$\Big(1+\sfrac{r C[dy,dx]-\eps}{n}\Big)^{n}\le
\Big(1+ \alpha q_{n}[y,x]\Big)^{n}\le \Big(1+\sfrac{r
C[dy,dx]+\eps}{n}\Big)^{n},$$ by the point-wise convergence. Hence by
the sandwich theorem and Euler's formula we get
(\ref{Ising.sequence}).
\end{Proof}

\begin{lemma}[Exponential tightness]\label{Ising.tightness}
For every $\Theta>0$  there exists  $M\in \N$ such that
$$\limsup_{n\rightarrow\infty}\sfrac{1}{n}\log P\Big\{|E|>n M \Big\}\le-\Theta.$$
\end{lemma}

\begin{Proof}
Let $c>\sup_{x,y\in\R} C[dy,dx]>0.$  Now, we  use a coupling argument to  define, for all large~$n$, a new spinned random
graph $\tilde{X}$ with spin law $\ell$ and bond probability
$\frac{d}{n}$, such that any bond present in $X$ is also present in
$\tilde{X}$. Let $|\tilde{E}|$ be the number of bonds of
$\tilde{X}$. Then,  by Chebyshev's inequality, the binomial distribution, and
Lemma~\ref{Ising.sequence1}, we have that
\begin{equation}\nonumber
\begin{aligned}
P\Big\{ |\tilde{E}|\ge n w\Big\}\le
e^{-nw}\me\big\{e^{|\tilde{E}|}\big\}&
 =e^{-nw}\sum_{k=0}^{\frac{n(n-1)}{2}}e^{k} \left(\heap{n(n-1)/2}{
 k}\right)\Big(\frac{d}{n}\Big)^{k}\Big(1-\frac{d}{n}\Big)^{n(n-1)/2-k}\\
&= e^{-nl}\Big( 1-\frac{d}{n}+e\frac{d}{n}\Big)^{n(n-1)/2} \le
e^{-nw}e^{nd(e-1+o(1))}.
\end{aligned}
\end{equation}
Now given $\theta>0$ choose $M\in\N$  such that $M> \Theta+d(e-1)$
and observe that, for extremely large $n,$
\begin{equation}\nonumber
P\big\{|E|\ge n M\big\}\le P\big\{ |\tilde{E}|\ge
nM\big\}\le e^{-n\Theta},
\end{equation}\nopagebreak
which implies the statement.\end{Proof}

\begin{lemma}[Exponential tightness]\label{Ising.tightness1}
For every $\Theta>0$  there exists  $K_{\Theta} \subset \skril(\R)$
such that
$$\limsup_{n\rightarrow\infty}\sfrac{1}{n}\log P\Big\{\skrim_G^1\not \in K_{\Theta}\Big\}\le-\Theta.$$
\end{lemma}
\begin{Proof}
Let  $w\in\N$  and  choose $k(w)\in\N,$ large enough,  such that
$\ell[e^{w^2\1_{\{s>k(w)\}}}]\le 2^{w}.$  Then, using  exponential Chebyschev's inequality, we have  that
$$P\Big\{\int_{\{x> k(w)\}}\skrim_G^1(dx)\ge \sfrac{1}{w}\Big\}\le
e^{-nw}\me \Big\{e^{w^2\sum_{u\in \skriv}\1_{\{\sigma(\rho(u))\ge
k(w)\}}}\Big\}\le e^{-nw}\big
(\ell[e^{w^2\1_{\{s>k(w)\}}}]\big)^{n}\le e^{-n(w-\log2)}.$$ Now we
fix $\Theta>0,$ choose $M>\theta+\log2,$   define the  set
$\Gamma_M$ by

$$ \Gamma_M:=\Big\{ \eta:\int_{\{x\ge
k(w)\}}\eta(dx)> \sfrac{1}{w}, \mbox{ for  all  $w\ge M $}\Big\}$$

As $\{x\le k(w)\}\subset\R$ is compact, the set $\Gamma_M$ is
pre-compact in the weak topology, by Prohorov's criterion. As
$$P\Big\{ \skrim_G^1 \not\in \Gamma_M \Big\}\le \frac{1}{1-e^{-1}} \exp(-n [M-\log2]),$$ we conclude that
$$\limsup_{n\to\infty} \frac 1n \log P\Big\{ \skrim_G^1\not\in K_\Theta
\Big\} \le -\Theta,$$ for the closure $K_\Theta$ of $\Gamma_M$ as
required for the proof.

\end{Proof}

\subsection{Proof of the upper bound in Theorem~\ref{main4}}\\

Denote by $\skric_1$ the space of  bounded functions on $\R$ and
by $\skric_2$ the space of bounded symmetric functions on
$\R\times\R$, and write
$$\hat{J}(\eta,\varpi):=\sup_{\heap{f\in\skric_{1}}{g\in\skric_{2}}}
\Big\{\int_{\R}\big(f[x]-U_{f}\big)\eta[dx]+\sfrac{1}{2}
\int_{\R\times\R}g[y,x]\varpi[dy,dx]+ \sfrac{1}{2}
\int_{\R\times\R}(1-e^{g[y,x]})C[dy,dx]\eta[dx]\eta[dy]\Big\}.$$

\begin{lemma}\label{Ising.L2L1uppbound}
For each closed set
$F\subset\skril(\R)\times\tilde{\skril}_*(\R\times\R),$
$$\limsup_{n\rightarrow\infty}\sfrac{1}{n}\log P\big\{(\skrim_G^1,\skrim_G^2)\in
F\big\}\le-\inf_{(\eta,\varpi)\in F}\hat{J}(\eta,\varpi).$$
\end{lemma}

\begin{Proof}
First let $\tilde{f}\in\skric_1$ and $\tilde{g}\in\skric_2$ be
arbitrary. Define $\tilde{\beta}\colon\R\times\R\rightarrow\R$ by
$$\tilde{\beta}[y,x]=(1-e^{\tilde{g}[y,x]})C[dy,dx].$$
Observe that, by Lemma~\ref{Ising.sequence1},
$\tilde{\beta}[y,x]=\lim_{n\rightarrow\infty}\tilde{h}_n[y,x]$ for
all $a,b\in\R$, recalling the definition of $\tilde{h}_n$ from
\eqref{hdef}. Hence, by (\ref{Ising.Itransform}), for sufficiently
large~$n$,
\begin{equation}\nonumber
e^{\sup_{x\in\R} | \tilde\beta(x,x)| } \ge\int
e^{\langle\frac{1}{2}\skrim_{\Delta}^{1},\,
\tilde{h}_n\rangle}d\tilde{ P} =\me\Big\{e^{n\langle \skrim_G^1,
\tilde{f}-U_{\tilde{f}}\rangle+n\langle\frac{1}{2}\skrim_G^2,
\tilde{g}\rangle+n\langle\frac{1}{2}\skrim_G^1\otimes \skrim_G^1,
\tilde{h}_n\rangle}\Big\},
\end{equation}
where $\skrim_{\Delta}^{1}=\sfrac{1}{n}\sum_{v\in \skriv}\delta_{(\sigma(\rho(v)),\sigma(\rho(v)))}$
and therefore,
\begin{equation}\label{Ising.estP}
\limsup_{n\rightarrow\infty}\sfrac{1}{n}\log\me\Big\{e^{n\langle
\skrim_G^1,\tilde{f}-U_{\tilde{f}}\rangle+n\langle \frac{1}{2}\skrim_G^2,
\tilde{g}\rangle+n\langle\frac{1}{2}\,\skrim_G^1\otimes \skrim_G^1 ,
\tilde{h}_n\rangle}\Big\}\le 0.
\end{equation}

Given $\eps>0$ let
$\hat{J}_{\eps}(\eta,\varpi)=\min\{\hat{J}(\eta,\varpi),{\eps}^{-1}\}-\eps.$
Suppose  that $(\eta,\varpi)\in F$ and observe that
$\hat{J}(\eta,\varpi)>\hat{J}_{\eps}(\eta,\varpi).$ We now fix
$\tilde{f}\in\skric_1$ and $\tilde{g}\in\skric_2$ such that
$$\langle\tilde{f}-U_{\tilde{f}},\eta\rangle+ \sfrac 12\,\langle
\tilde{g},\varpi\rangle+ \sfrac 12 \, \langle
\tilde{\beta},\eta\otimes\eta\rangle\ge\hat{J}_{\eps}(\eta,\varpi)
.$$ As $\tilde{f}, \tilde{g}$ are  bounded functions, there exist
open neighbourhoods $B_{\varpi}^{2}$ and $B_{\eta}^{1}$ of
$\varpi,\eta$ such that
\begin{equation}\nonumber
\inf_{\heap{\tilde{\eta}\in B_{\eta}^{1}}{\tilde{\varpi}\in
B_{\varpi}^{2}}}\big\{\langle \tilde{f}-U_{\tilde{f}} ,
\tome\rangle+ \sfrac12 \, \langle
\tilde{g},\tilde\varpi\rangle+\sfrac 12 \, \langle
\tilde{\beta},\tome\otimes\tome\rangle\big\}\ge
\hat{J}_{\eps}(\eta,\varpi)-\eps.
\end{equation}
Using Chebyshev's inequality and (\ref{Ising.estP}) we have that
\begin{equation}\begin{aligned}
\limsup_{n\rightarrow\infty}\sfrac{1}{n}& \log P\big\{(\skrim_G^1,\skrim_G^2)\in
B_{\eta}^{1}\times B_{\varpi}^{2}\big\}\\
&\le\limsup_{n\rightarrow\infty}\sfrac{1}{n}\log
\me\Big\{e^{n\langle \skrim_G^1,\tilde{f}-U_{\tilde{f}}\rangle+n\langle
\frac{1}{2}\skrim_G^2,\tilde{g}\rangle+n\langle\frac{1}{2}\skrim_G^1\otimes
\skrim_G^1,\tilde{h}_n\rangle}\Big\}-\hat{J}_{\eps}(\eta,\varpi)+\eps\\
&\le -\hat{J}_{\eps}(\eta,\varpi)+\eps.\label{Ising.LDPballs}
\end{aligned}\end{equation}
Now we use Lemma~\ref{Ising.tightness}  and Lemma~\ref{Ising.tightness1} with $\Theta=\eps^{-1},$ to
choose $M(\eps)\in\N$  and  $K_{\eps}$ such that
\begin{equation}\label{unboundedset}
\limsup_{n\rightarrow\infty}\sfrac{1}{n}\log P\Big\{|E|> n
M(\eps) \Big\}\le-\eps^{-1}  \,\mbox{ and}\,
 \limsup_{n\rightarrow\infty}\sfrac{1}{n}\log P\Big\{\skrim_G^1\not\in K_{\eps}\Big\}\le-\eps^{-1}  \,
\end{equation}
For this $M(\eps)$  and  $K_{\eps}$ we define the set
$\Gamma_{\eps}$   by
$$\Gamma_{\eps}=\Big\{(\eta,\varpi)\in\skril(\R)\times\tilde{\skril}_*(\R\times\R):\eta\in K_{\eps}^{c},\,\|\varpi\|\le
2M(\eps)\Big\},$$ and recall that $\|\skrim_G^2\|=2{|E|}/{n}.$ The set
$\Gamma_{\eps}\cap F$ is compact and hence we  can  cover it  by
finitely many sets $B_{\eta_ {r}}^{1}\times B_{\varpi_
{r}}^{2},r=1,\ldots ,m$ with $ (\eta_{r},\varpi_{r})\in F$ for
$r=1,\ldots ,m.$ Consequently,
\begin{equation}\nonumber
 P\big\{(\skrim_G^1,\skrim_G^2)\in F\big\}\le\sum_{r=1}^{m} P\big\{(\skrim_G^1,\skrim_G^2)\in B_{\eta_{r}}^{1}\times
B_{\varpi_{r}}^{2}\big\}+ P\big\{(\skrim_G^1,\skrim_G^2)\not\in
\Gamma_{\eps}\big\}.
\end{equation}
We may now use (\ref{Ising.LDPballs}) and \eqref{unboundedset} to
obtain, for all small $\eps>0$,
\begin{equation}\nonumber
\begin{aligned}
\limsup_{n\rightarrow\infty}\sfrac{1}{n}\log P\big\{(\skrim_G^1,\skrim_G^2)\in
F\big\}& \le \max_{r=1}^{m} \Big(
\limsup_{n\rightarrow\infty}\sfrac{1}{n}\log P\big\{(\skrim_G^1,\skrim_G^2)\in
B_{\eta_{r}}^{1}\times B_{\varpi_{r}}^{2}\big\} \Big) \vee (-\eps)^{-1} \\
& \le \Big( -\inf_{(\eta,\varpi)\in
F}\hat{J}_{\eps}(\eta,\varpi)+\eps \Big)\vee (-\eps)^{-1}.
\end{aligned}
\end{equation}
Taking $\eps\downarrow 0$ we get the desired statement.
\end{Proof}

Moreover, we  can write the rate function in terms of relative entropies,
see \cite[(2.15)]{21}, and hence we will  show that the  rate function
is a good rate function. Recall the definition of the function $J$
from Theorem~\ref{main4}.

\begin{lemma}\label{Ising.Vrate}\ \\ \vspace{-.6cm}
\begin{itemize}
\item[(i)] $\hat{J}(\eta,\varpi)=J(\eta,\varpi),$ for any
$(\eta,\varpi)\in\skril(\R)\times\tilde{\skril}_*(\R\times\R)$,
\item[(ii)] $J$ is a good rate function and
\item[(iii)] $\Big[H\big(\varpi\,\|\,C\eta\otimes\eta\big)+\|
C\eta\otimes\eta \| -\|\varpi\|\Big]\ge 0$ with equality iff
$\varpi=C\eta\otimes\eta.$
\end{itemize}

\end{lemma}
\begin{Proof}
(i) If $\varpi\not\ll C\eta\otimes\eta,$ then, there
exists
 $y_0,x_0 \in\R$ with $C\eta\otimes\eta(y_0,x_0)=0$ and
 $\varpi(y_0,x_0)>0.$ We define $\hat{g}\colon \R\times\R\rightarrow \R$  by
$$ \hat{g}[y,x]=\log\big[K(\one_{(y_0,x_0)}[y,x]+\one_{(y_0,x_0)}[y,x])+1\big], \mbox{ for
$y,x\in\R$ and  $K>0.$ }$$ This choice of $\hat{g}$ and $f=0$  gives

\begin{equation}\nonumber
\begin{aligned}
&\int_{\R}\big(f[x]-U_{f}\big)\eta(dx)+\int_{\R\times\R} \sfrac 12
\hat{g}[y,x]\varpi[dy,dx]+\int_{\R\times\R}\sfrac 12(1-e^{\hat{g}[y,x]})C[dy,dx]\eta[dx]\eta[dy]\\
&\ge \sfrac 12 \log(K+1)\varpi(y_{0},x_{0}) \to \infty, \qquad
\mbox{ for $K\uparrow\infty.$ }
\end{aligned}
\end{equation}
Now suppose that $\varpi\ll C\eta\otimes\eta.$ We have
\begin{equation}\nonumber
\begin{aligned}
\hat{J}(\eta,\varpi) & = \sup_{f\in\skric_{1}}
\Big\{\int_{\R}\Big(f[x]-\log\int_{\R}e^{{f}[x]}\ell[dx]\Big)
\, \eta[dx] \Big\} \\
&\qquad  + \sfrac{1}{2} \int_{\R\times\R}
C[dy,dx]\eta[dx]\eta[dy] +  \sfrac{1}{2} \sup_{g\in\skric_{2}}
\Big\{ \int_{\R\times\R}g[y,x]\varpi[dy,dx]- \int_{\R\times\R}
e^{g[y,x]} C[dy,dx]\eta[dx]\eta[dy]\Big\}.
\end{aligned}
\end{equation}
By the variational characterization of relative entropy, the first
term equals $H(\eta \, \| \, \ell)$. By the substitution $h=e^{g}
\, \frac{C \eta \otimes \eta}{\varpi}$ the last term equals
\begin{equation}\nonumber
\begin{aligned}
\sup_{\heap{h\in\skric_{2}}{h \ge 0}} & \int_{\R\times\R} \Big[ \log
\Big( h[y,x] \frac{\varpi[dy,dx]}{C[dy,dx]\eta[dx]
\eta[dy]}\Big)
-h[y,x] \Big]  \, \varpi[dy,dx] \\
& = \sup_{\heap{h\in\skric_{2}}{h \ge 0}} \sum _{x,y\in\R} \big(
\log h[y,x] - h[y,x] \big) \, \varpi[dy,dx] +
\sum _{x,y\in\R} \log \Big( \frac{\varpi[dy,dx]}{C[dy,dx] \eta[dx] \eta[dy]} \Big)\, \varpi[dy,dx] \\
& = - \| \varpi\|  + H(\varpi \, \| \, C \eta \otimes \eta ),
\end{aligned}
\end{equation}
where we have used $\sup_{z>0} \log z - z = -1$ in the last line.
This gives that $\hat{J}(\eta,\varpi)={J}(\eta,\varpi)$.

(ii) We write
${h_C}(\varpi\,\|\,\eta):=\Big[\, H\big(\varpi\,\|\,C\eta\otimes\eta\big)+\, \|
C\eta\otimes\eta \| -\, \|\varpi\|\Big]$  and notice that,  all summands are
continuous in $\eta, \varpi.$   Thus $I$ is a rate function.
Moreover, for all $r<\infty$, the level sets $\{J(\eta,\varpi) \le r\}$
are contained in the bounded set
$\{(\eta,\varpi)\in\skril(\R)\times\tilde{\skril}_*(\R\times\R)\colon
\,{h_C}(\varpi\,\|\,\eta)\le r\}$ and are therefore
compact. Consequently, $I$ is a good rate function.

(iii)  We consider the non-negative function $\phi(z)=z\log z-z+1$, for
$z> 0$, $\phi(0)=1$, which has its only root in $z=1$. Note that
\begin{align}\label{Ising.calculusnot}
{h_C}(\varpi\,\|\,\eta)=
\left\{\begin{array}{ll}\int\phi\circ g\,\,dC\eta\otimes \eta
&\mbox{ if
$g:=\sfrac{d\varpi}{dC\eta\otimes\eta}\ge 0$  exists, }\\
\infty & \mbox{ otherwise.}
\end{array}\right.
\end{align}
Hence ${h_C}(\varpi\,\|\,\eta)\ge 0$, and, if
$\varpi=C\eta\otimes\eta,$ then
$\phi(\sfrac{d\varpi}{dC\eta\otimes\eta})=\phi(1)=0$ and so
${h_C}(C\eta\otimes\eta\,\|\,\eta)=0$. Conversely,
if ${h_C}(\varpi\,\|\,\eta)=0$, then $\varpi[dy,dx]>0$
implies $C\eta\otimes\eta[y,x]>0$, which then gives $\phi\circ
g[y,x]=0$ and  $g[y,x]=1$. Therefore, $\varpi=C\eta\otimes
\eta,$ which ends the proof of (iii) above.
\end{Proof}


\subsection{Proof of the lower bound in Theorem~\ref{main4} }\\

\begin{lemma}\label{Ising.lowbound1}
For every open set
$O\subset\skril(\R)\times\tilde{\skril}_*(\R\times\R),$
$$\liminf_{n\rightarrow\infty}\sfrac{1}{n}\log P\Big\{(\skrim_G^1,\skrim_G^2)\in
O\Big\}\ge -\inf_{(\eta,\varpi)\in O}J(\eta,\varpi).$$
\end{lemma}

\begin{Proof}
Suppose  we have $(\eta,\varpi)\in O,$ with  $\varpi\ll
C\eta\otimes\eta$. Define $\tilde{f}_{\eta}\colon\R\rightarrow
\R$  by
\begin{equation}\label{Ising.S1}
\begin{aligned}\nonumber
\tilde{f}_{\eta}(a)=\left\{\begin{array}{ll}\log\sfrac{\eta[dx]}{\ell[dx]},
&\mbox{if $\eta[dx]> 0$,  }\\
0, & \mbox{otherwise.}
\end{array}\right\}.
\end{aligned}
\end{equation}
and $\tilde{g}_{\varpi}\colon\R\times\R\rightarrow \R$ by
\begin{equation}\label{Ising.S2}
\begin{aligned}\nonumber
\tilde{g}_{\varpi}[y,x]=\left\{\begin{array}{ll}\log\sfrac{\varpi[dy,dx]}{C[dy,dx]\eta[dx]\eta[dy]},&\mbox{if $\varpi[dy,dx]>0$, }\\
0, & \mbox{otherwise.}
\end{array}\right.
\end{aligned}
\end{equation}

In addition, we let
$\tilde{\beta}_{\varpi}[y,x]=C[dy,dx](1-e^{{\tilde
g}_{\varpi}[y,x]})$ and note that
$\tilde{\beta}_{\varpi}[y,x]=\lim_{n\rightarrow\infty}\tilde{h}_{\varpi,
n}[y,x],$ for all $x,y\in\R$  where
\begin{equation}\nonumber
\tilde{h}_{\varpi,
n}[y,x]=\log\Big[\big(1-q_n[dy,dx]+q_n[dy,dx]e^{\tilde{g}_{\varpi}[y,x]}\big)^{-n}\Big].
\end{equation}
Choose $B_{\eta}^{1},B_{\varpi}^{2}$  open neighbourhoods of
$\eta,\varpi,$ such that  $B_{\eta}^{1}\times
B_{\varpi}^{2}\subset O$ and for all
$(\tilde{\eta},\tilde{\varpi})\in B_{\eta}^{1}\times
B_{\varpi}^{2}$
$$\langle \tilde{f}_{\eta},\eta\rangle+\sfrac 12\, \langle\tilde{g}_{\varpi},\varpi\rangle+
\sfrac 12\,
\langle\tilde{\beta}_{\varpi},\eta\otimes\eta\rangle-\eps\le
\langle \tilde{f}_{\eta},\tome\rangle+ \sfrac 12\, \langle
\tilde{g}_{\varpi},\tilde\varpi\rangle+ \sfrac 12\,
\langle\tilde{\beta}_{\varpi},\tome\otimes\tome\rangle.$$

 We now use
$\tilde{ P},$ the  probability distribution obtained by transforming
$ P$ using the functions $\tilde{f}_{\eta}$,
$\tilde{g}_{\varpi}$. Note that the spin law in the transformed
distribution is now $\eta$, and the connection
 probabilities~$\tilde{q}_n[dy,dx]$ satisfy $n \, \tilde{q}_n[dy,dx]
\to (\varpi[dy,dx])/(\eta[dx]\eta[dy]) =: \tilde C[dy,dx],$
 as  $n\to\infty.$ We use (\ref{Ising.Itransform}), to obtain
\begin{equation}
\begin{aligned}
P\Big\{(\skrim_G^1,\skrim_G^2)& \in
O\Big\}\ge\tilde{\me}\Big\{\sfrac{d P}{d\tilde{ P}}(G)\one_{\{(\skrim_G^1,\skrim_G^2)\in
B_{\eta}^{1}\times B_{\varpi}^{2}\}}\Big\}\\
&=\tilde{\me}\Big\{\prod_{u\in V}e^{-\tilde{f}_{\eta}[\sigma(\rho(u))]}\prod_{(v,u)\in
E}e^{-\tilde{g}_\varpi[\sigma(\rho(v)),\sigma(\rho(u))]}\prod_{(v,u)\in
\skrie}e^{-\sfrac 1n \, \tilde{h}_{\varpi, n}[\sigma(\rho(v)),\sigma(\rho(u))]}
\one_{\{(\skrim_G^1,\skrim_G^2)\in B_{\eta}^{1}\times B_{\varpi}^{2}\}}\Big\}\\
&=\tilde{\me}\Big\{ e^{-n\langle L^1,
\tilde{f}_{\eta}\rangle-n\frac{1}{2}\, \langle L^2,
\tilde{g}_{\varpi}\rangle-n\frac{1}{2}\,\langle \skrim_G^1\otimes
\skrim_G^2, {\tilde{g}}_{\varpi}\rangle+\frac{1}{2}\,\langle \skrim_{\Delta}^{1},\tilde{h}_{\varpi, n}\rangle}\times\one_{\{(\skrim_G^1,\skrim_G^2)\in B_{\eta}^{1}\times B_{\varpi}^{2}\}}\Big\}\\
&\ge \exp\big(-n\langle \tilde{f}_{\eta},\eta\rangle-n\sfrac 12
\langle\tilde{g}_{\varpi},\varpi\rangle-n \sfrac 12
\langle\tilde{\beta}_\varpi,\eta\otimes\eta\rangle
+m-n\eps\big)\times\tilde{ P}\Big\{(\skrim_G^1,\skrim_G^2)\in
B_{\eta}^{1}\times B_{\varpi}^{2}\Big\},
\end{aligned}
\end{equation}
where  $m:=0 \wedge \inf_{x\in\R}\tilde\beta[x,x].$ Therefore, by
(\ref{Ising.sequence}), we have
\begin{align}\label{Ising.lowbound2}
&\liminf_{n\rightarrow\infty}\sfrac{1}{n}\log P\Big\{(\skrim_G^1,\skrim_G^2)\in
O\Big\}\nonumber\\
&\ge-\langle \tilde{f}_{\eta},\eta\rangle-\sfrac12 \,
\langle\tilde{g}_{\varpi},\varpi\rangle-\sfrac 12\,
\langle{\tilde{{\beta}}_{\varpi}},
\eta\otimes\eta\rangle-\eps+\liminf_{n\rightarrow\infty}\sfrac{1}{n}\log\tilde{ P}
\Big\{(\skrim_G^1,\skrim_G^2)\in B_{\eta}^{1}\times
B_{\varpi}^{2}\Big\}.\nonumber
\end{align}

The result follows once we prove that
\begin{equation}\label{Ising.translowbound}
\liminf_{n\rightarrow\infty}\sfrac{1}{n}\log\tilde{ P}\Big\{(\skrim_G^1,\skrim_G^2)\in
B_{\eta}^{1}\times B_{\varpi}^{2}\Big\}= 0.
\end{equation}
We apply the upper bound (but now with $\tilde{ P}$ in  place  of  $ P$) to prove (\ref{Ising.translowbound}). Then we
obtain
\begin{equation}
\begin{aligned}
\limsup_{n\rightarrow\infty}\sfrac{1}{n}\log\tilde P\big\{(\skrim_G^1,\skrim_G^2)
\in (B_{\eta}^{1}\times B_{\varpi}^{2})^{c} \big\}&\le
-\inf_{(\tilde{\eta},\tilde{\varpi})\in \tilde{F}}
\tilde{J}(\tilde{\eta},\tilde\varpi),
\end{aligned}
\end{equation}
where $\tilde{F}=(B_{\eta}^{1}\times B_{\varpi}^{2})^{c}$ and
$\tilde{J}(\tilde\eta, \tilde\varpi):= H(\tilde\eta \, \| \,
\eta) + \sfrac 12 {h_{\tilde C}}(\tilde\varpi\,\|\,
\tilde\eta)$. It therefore reduces to showing  that the infimum is
non-negative. Suppose that by  contradiction we  have  a sequence
$(\tilde{\eta}_n,\tilde{\varpi}_n)\in\tilde{F}$ with
$\tilde{J}(\tilde{\eta}_n,\tilde\varpi_n)\downarrow 0.$ Then, as
$\tilde{J}$ is a good rate function and its level sets are compact,
and by  lower semi-continuity of the map
$({\tilde{\eta}},\tilde{\varpi})\mapsto\tilde{J}({\tilde{\eta}},\tilde{\varpi})$,
we can find a limit point $(\tilde{\eta},\tilde\varpi)\in\tilde{F}$
with $\tilde{J}(\tilde{\eta},\tilde{\varpi})=0$ . By
Lemma~\ref{Ising.Vrate} this implies $H(\tilde{\eta}\,\|\,\eta)=0$ and
${h_C}(\tilde{\varpi}\,\|\,\tilde{\eta})=0$, hence
$\tilde{\eta}=\eta,$ and
$\tilde{\varpi}=\tilde{C}\tilde{\eta}\otimes\tilde{\eta}=\varpi$ contradicting
$(\tilde{\eta},\tilde{\varpi})\in\tilde{F}$.
\end{Proof}

\section{Proof  of  Main  Results}\label{mainproof}
\subsection{Proof  of  Theorem~\ref{mainpart}}
We  kick  start the proof of  our  Thermodynamics limit results  by
stating an  important Lemma (Varadhan's Lemma,  see
\cite[Theorem~4.3.1]{21} ), which is a key step in establishing
our first result Theorem~\ref{main1}, without   proof.
\begin{lemma}[{\bf Varadhan}]\label{Ising.Varad}
Suppose  the  functional  $M_n$  from the space  of  finite spinned
graphs to  $\skril$  satisfies  the  LDP  withh  good  rate function
$J:\skril \to  [0,\infty]$    and let   $\Psi:\skril \to \R$ be  any
continuous  function. Assume  further  the  following  moment
condition   for  some  $\lambda>1$,
$$ \limsup_{n\to\infty}\sfrac{1}{n}\log\me\big[e^{n\lambda
\Psi(M_n[X])}\big]<\infty .$$ Then,
$$ \lim_{n\to\infty}\sfrac{1}{n}\log\me\big[e^{n\lambda
\Psi(M_n[X])}\big]=\sup_{m\in\skril}\big\{\Psi(m)-J(m)\big\} .$$
\end{lemma}
Now we prove an annealed asymptotic result  for  the  partition
function  for the  ferromagnetic Ising  model on  spinned  random
graphs,  as the  graph  size goes to  infinity.

\begin{Proof}  Recall  the partition function of  the Ising
model  on the spinned random graph network from section~\ref{section2} and write it
as integration  of  some function with respect to our  empirical
distributions:
\begin{equation}
 {\sf
E}[Z_G(B,\beta)]:=2^n\me\Big[\exp\Big\{\sfrac{n}{2}\int\sfrac{\sigma(x)}{B(x)}\sfrac{\sigma(x)}{B(y)}\skrim_G^2[d\sigma(x),d\sigma(y)]
+n\int\sfrac{\sigma(x)}{\sqrt{\beta}}\skrim_G^1[d\sigma(x)]\Big\}\Big]
\end{equation}
Now  using the (Varadhan) Lemma~\ref{Ising.Varad}  and
Theorem~\ref{main4} we  obtain,

\begin{equation}\begin{aligned}\label{Ising.happy}
\lim_{n\to\infty}  \, \sfrac 1n \log {\sf E} [Z_G(B,\beta)] &=\log2+
\sup \Big\{ \sfrac 1 2\, \int
\sfrac{\sigma(x)}{B(x)}\sfrac{\sigma(x)}{B(y)}\varpi[d\sigma(x),\,d\sigma(y)]+\int\sfrac{\sigma(x)}{\sqrt{\beta}}\eta[d\sigma(x)]-J(\eta,\varpi)\\
&\qquad\qquad\qquad:\eta\in \skrix(\{\sigma(-1),\sigma(+1)\}), \varpi\in
\skril_*(\{\sigma(-1),\sigma(+1)\}\times
\{\sigma(-1),\sigma(+1)\})\Big\}\\
 & = \sup \Big\{ \sfrac \beta 2\, \big(
\varpi(\Delta) -
\varpi(\Delta^{\rm c}) \big)+B(1)t-B(-1)(1-t) -t\log(t)-(1-t)\log(1-t)\\
&\qquad\qquad - \sfrac 12 \big( H(\varpi\,\|\, \eta_t) +
C_{+1,+1}t+C_{-1,-1}(1-t)+2C_{-1,+1}t(1-t) - \|\varpi\| \big) \Big\}\,
,
\end{aligned}
\end{equation}
where $\Delta$ is the diagonal in $\{\sigma(-1),\sigma(+1)\} \times
\{\sigma(-1),\sigma(+1)\}$, and the supremum is over all $t\in[0,1]$ and
$\varpi\in \skril_*(\{\sigma(-1),\sigma(+1)\} \times \{\sigma(-1),\sigma(+1)\})$, and the
distribution $\eta_t\in\tilde\skril_*(\{\sigma(-1),\sigma(+1)\}\times
\{\sigma(-1),\sigma(+1)\})$ is defined by
$$\eta_t[s(i),s(j)]=C_{i,j}t^{(2+i+j)/2}(1-t)^{(2-i-j)/2}\, \mbox{ for }
i,j\in\{-1,+1\}\, .$$

We take  the partial derivatives of  \eqref{Ising.happy}  with
respect  to $\varpi,$    and  set our results to  zero to  get,
$$\varpi[\sigma(+1),\sigma(+1)]= e^{\beta}\eta_t[\sigma(+1),\sigma(+1)],$$
$$\varpi[\sigma(-1),\sigma(-1)]= e^{\beta}\eta_t[\sigma(-1),\sigma(-1)]$$
$$\varpi[\sigma(+1),\sigma(-1)]= e^{-\beta}\eta_t[\sigma(+1),\sigma(-1)]$$
$$\varpi[\sigma(-1),\sigma(+1)]= e^{-\beta}\eta_t[\sigma(-1),\sigma(+1)].$$

Therefore  writing  $\varpi$  in \eqref{Ising.happy}  we obtain

$$
\begin{aligned}
&\lim_{n\to\infty}  \, \sfrac{1}{n} \log {\sf E} [Z_G(B,\beta)] \\
&=\sup_{t\in[0,1]} \Big\{ \, \sfrac{\beta}{2}e^{\beta}\eta_t[\sigma(+1),\sigma(+1)]+ \sfrac{\beta}{2}e^{\beta}\eta_t[\sigma(-1),\sigma(-1)]
- \beta e^{-\beta}\eta_t[\sigma(-1),\sigma(+1)]+B(1)x-B(-1)(1-t)\\
&-t\log(t)-(1-t)\log(1-t) -  \sfrac{\beta}{2} e^{\beta}\eta_t[\sigma(+1),\sigma(+1)]- \sfrac{\beta}{2}e^{\beta}\eta_t[\sigma(-1),\sigma(-1)]+\beta e^{-\beta}\eta_t[\sigma(-1),\sigma(+1)] -\sfrac {1}{2} C_{+1,+1}t\\
&-\sfrac {1}{2} C_{-1,-1}(1-t)-C_{-1,+1}t(1-t)  + \sfrac {1}{2} e^{\beta}\eta_t[\sigma(+1),\sigma(+1)]+ \sfrac{1}{2} e^{\beta}\eta_t[\sigma(-1),\sigma(-1)]+e^{-\beta}\eta_t[\sigma(-1),\sigma(+1)] \Big\}\\
& = \sup_{t\in[0,1]} \Big\{ \, -t\log(t)-(1-t)\log(1-t)  -\sfrac {1}{2}C_{+1,+1}t^2 -\sfrac 12 C_{-1,-1}(1-t)^2+B(1)x-B(-1)(1-t)
&-C_{-1,+1}t(1-t)  + \sfrac {1}{2} e^{\beta}C_{+1,+1}t^2 + \sfrac {1}{2}e^{\beta}C_{-1,-1}(1-t)^2+e^{-\beta}C_{-1,+1}t(1-t) \Big\}\\
& = \sup_{t\in[0,1]} \Big\{ \, -t\log(t)-(1-t)\log(1-t)  +\sfrac {1}{2} C_{+1,+1}(e^{\beta}-1)x^2+\sfrac {1}{2} C_{-1,-1}(e^{\beta}-1)(1-t)^2
+B(1)t-B(-1)(1-t) +C_{-1,+1}(e^{-\beta}-1)x(1-t) \Big\}\\
& = \sup_{t\in[0,1]} \Big\{ \, -t \log t-(1-t)\log(1-t)+a_1t^2+ a_2(1-t)^2+a_3t(1-t)+B(1)t-B(-1)(1-t)\Big\}\\
& = \sup_{t\in[0,1]} \Big\{ \, -t \log t-(1-t)\log(1-t)+\lambda(t)\Big\}
\end{aligned}$$

which  ends  the proof  of  our  main  Theorem.

\end{Proof}
\subsection{Proof  of  Corollary~\ref{mainpart1}}.
Note  that  in case  of  the  Erdos-Renyi  graphs  $C_{i,j}=c$   for  all  $i,j=-1,\,1,$ with  $np_n\to c$ and $B(1)=B(-1)=b.$ Therefore  we  can  invoke   Theorem~\ref{mainpart}  to obtain   \eqref{mainErdos-Renyi} as required. This  ends  the  proof.

\section{Summary, Discussions  and  Future Work}
In  this  article  we  have  found  annealed  asymptotic  result  for  the random  partition  function  of  the  Ising  model on  inhomogeneous graphs. The  main  technique used  to  establish  the result  is  joint  LDP  for  suitably  defined  empirical  measures  of   the  spinned  random  graph. Thus,  we  defined  for each  spinned  random  graph  an  empirical  spin measure  and  empirical  bond  measure. For  these  empirical  measures, the  joint  LDP  of  \cite{17}  is  extended to cover   continuous   spins space.  Note  that  the  hamiltonian  of  the  Ising  model   was  then  written  as  the  integral  of  a  function  and  the  empirical  measures. The Varadhan's Lemma,  see
\cite[Theorem~4.3.1]{21}   is  applied  to the  LDP  to establish the annealed asymptotic result  for  normalized  log expected  value  of  the  random  partition  function  and  the  main result was  obtained as  a  solution  of  the  optimization  problem \ref{Ising.happy}. Elementary  calculus  gave  us  the  solution  of   \ref{Ising.happy}  which  ends  the  proof  of  the  main  results.\\

Corollary~\ref{mainpart1}  above  is  qualitatively  different  from  \cite[Theorem~5.1]{22}  in  two  folds:  First  the  result  in  this  paper  is  an  annealed  results  for  the  log-partition  of  the Erdos-Renyi  graphs  while \cite[Theorem~5.1]{22}  gives a  quenched  result  for  the  log-partition  function.  Second,  finding  explicit  value  for  the  normalized  log-partition in  \cite[Theorem~5.1]{22}  will  involve the  computation  of  expectations with  respect  to  the  law  of  a  local  tree  and   might  be   very difficult   without  a numerical  integration. However,  Corollary~\ref{mainpart1} of  this  paper presents  the  limit  of  the  normalized   log- expected  partition function as  a  simple  closed  optimization  problem.\\

The  same  method  may  be  adapted  for  random graphs  mentioned  in  the  Thesis  of  Dommers~\cite[Page~30]{22}  but  not  studied as  their  local  structure  is  not tree-like. i.e  conﬁguration model with a household structure  and scale-free percolation clusters  and  the  fitness preferential  attachment model. Combinatoric  arguments  through  the  method  of  types  similar  to  the  one  deployed  in  the  paper  \cite{19}  will  give  the  LDP.


\centerline{\textit{University of Ghana}}
\centerline{\textit{Statistics  Department, Box LG 115, Legon, Ghana}}
\centerline{\textit{Email: kdoku-amponsah@ug.edu.gh}}

\bigskip
\end{document}